
\def\singlespace{\normalbaselines}
\def\oneandahalfspace{\baselineskip=1.15\normalbaselineskip plus 1pt
\lineskip=2pt\lineskiplimit=1pt}

\def\np{\vfill\eject}
\def\nl{\hfil\break}

\def\nofirstpagenoten{\nopagenumbers\footline={\ifnum\pageno>1\tenrm
\hss\folio\hss\fi}}
\def\nofirstpagenotwelve{\nopagenumbers\footline={\ifnum\pageno>1\twelverm
\hss\folio\hss\fi}}
\def\leaderfill{\leaders\hbox to 1em{\hss.\hss}\hfill}
\def\ft#1#2{{\textstyle{{#1}\over{#2}}}}
\def\frac#1/#2{\leavevmode\kern.1em
\raise.5ex\hbox{\the\scriptfont0 #1}\kern-.1em/\kern-.15em
\lower.25ex\hbox{\the\scriptfont0 #2}}
\def\sfrac#1/#2{\leavevmode\kern.1em
\raise.5ex\hbox{\the\scriptscriptfont0 #1}\kern-.1em/\kern-.15em
\lower.25ex\hbox{\the\scriptscriptfont0 #2}}


\parindent=20pt
\def\narrow{\advance\leftskip by 40pt \advance\rightskip by 40pt}

\def\AB{\bigskip
        \centerline{\bf ABSTRACT}\medskip\narrow}
\def\nonarrower{\advance\leftskip by -40pt\advance\rightskip by -40pt}
\def\AE{\bigskip\nonarrower}

\def\boxit#1{\vbox{\hrule\hbox{\vrule\kern3pt
        \vbox{\kern3pt#1\kern3pt}\kern3pt\vrule}\hrule}}

\def\gtorder{\mathrel{\raise.3ex\hbox{$>$}\mkern-14mu
             \lower0.6ex\hbox{$\sim$}}}
\def\ltorder{\mathrel{\raise.3ex\hbox{$<$}|mkern-14mu
             \lower0.6ex\hbox{\sim$}}}
\def\dalemb#1#2{{\vbox{\hrule height .#2pt
        \hbox{\vrule width.#2pt height#1pt \kern#1pt
                \vrule width.#2pt}
        \hrule height.#2pt}}}

\font\fourteentt=cmtt10 scaled \magstep2
\font\fourteenbf=cmbx12 scaled \magstep1
\font\fourteenrm=cmr12 scaled \magstep1
\font\fourteeni=cmmi12 scaled \magstep1
\font\fourteenss=cmss12 scaled \magstep1
\font\fourteensy=cmsy10 scaled \magstep2
\font\fourteensl=cmsl12 scaled \magstep1
\font\fourteenex=cmex10 scaled \magstep2
\font\fourteenit=cmti12 scaled \magstep1
\font\twelvett=cmtt10 scaled \magstep1 \font\twelvebf=cmbx12
\font\twelverm=cmr12 \font\twelvei=cmmi12
\font\twelvess=cmss12 \font\twelvesy=cmsy10 scaled \magstep1
\font\twelvesl=cmsl12 \font\twelveex=cmex10 scaled \magstep1
\font\twelveit=cmti12
\font\tenss=cmss10
 
 \font\ninebf=cmbx7 scaled \magstep1
\font\ninerm=cmr7 scaled \magstep1 \font\ninei=cmmi7 scaled \magstep1
\font\ninesy=cmsy7 scaled \magstep1 
\font\eightrm=cmr7 scaled 1140 
 
\font\sevenbf=cmbx7 \font\sevenrm=cmr7 \font\seveni=cmmi7
\font\sevensy=cmsy7 

\catcode`@=11
\newskip\ttglue
\newfam\ssfam

\def\fourteenpoint{\def\rm{\fam0\fourteenrm}
\textfont0=\fourteenrm \scriptfont0=\tenrm \scriptscriptfont0=\sevenrm
\textfont1=\fourteeni \scriptfont1=\teni \scriptscriptfont1=\seveni
\textfont2=\fourteensy \scriptfont2=\tensy \scriptscriptfont2=\sevensy
\textfont3=\fourteenex \scriptfont3=\fourteenex \scriptscriptfont3=\fourteenex
\def\it{\fam\itfam\fourteenit} \textfont\itfam=\fourteenit
\def\sl{\fam\slfam\fourteensl} \textfont\slfam=\fourteensl
\def\bf{\fam\bffam\fourteenbf} \textfont\bffam=\fourteenbf
\scriptfont\bffam=\tenbf \scriptscriptfont\bffam=\sevenbf
\def\tt{\fam\ttfam\fourteentt} \textfont\ttfam=\fourteentt
\def\ss{\fam\ssfam\fourteenss} \textfont\ssfam=\fourteenss
\tt \ttglue=.5em plus .25em minus .15em
\normalbaselineskip=16pt
\abovedisplayskip=16pt plus 4pt minus 12pt
\belowdisplayskip=16pt plus 4pt minus 12pt
\abovedisplayshortskip=0pt plus 4pt
\belowdisplayshortskip=9pt plus 4pt minus 6pt
\parskip=5pt plus 1.5pt
\setbox\strutbox=\hbox{\vrule height12pt depth5pt width0pt}
\let\sc=\tenrm
\let\big=\fourteenbig \normalbaselines\rm}
\def\fourteenbig#1{{\hbox{$\left#1\vbox to12pt{}\right.\n@space$}}}

\def\twelvepoint{\def\rm{\fam0\twelverm}
\textfont0=\twelverm \scriptfont0=\ninerm \scriptscriptfont0=\sevenrm
\textfont1=\twelvei \scriptfont1=\ninei \scriptscriptfont1=\seveni
\textfont2=\twelvesy \scriptfont2=\ninesy \scriptscriptfont2=\sevensy
\textfont3=\twelveex \scriptfont3=\twelveex \scriptscriptfont3=\twelveex
\def\it{\fam\itfam\twelveit} \textfont\itfam=\twelveit
\def\sl{\fam\slfam\twelvesl} \textfont\slfam=\twelvesl
\def\bf{\fam\bffam\twelvebf} \textfont\bffam=\twelvebf
\scriptfont\bffam=\ninebf \scriptscriptfont\bffam=\sevenbf
\def\tt{\fam\ttfam\twelvett} \textfont\ttfam=\twelvett
\def\ss{\fam\ssfam\twelvess} \textfont\ssfam=\twelvess
\tt \ttglue=.5em plus .25em minus .15em
\normalbaselineskip=14pt
\abovedisplayskip=14pt plus 3pt minus 10pt
\belowdisplayskip=14pt plus 3pt minus 10pt
\abovedisplayshortskip=0pt plus 3pt
\belowdisplayshortskip=8pt plus 3pt minus 5pt
\parskip=3pt plus 1.5pt
\setbox\strutbox=\hbox{\vrule height10pt depth4pt width0pt}
\let\sc=\ninerm
\let\big=\twelvebig \normalbaselines\rm}
\def\twelvebig#1{{\hbox{$\left#1\vbox to10pt{}\right.\n@space$}}}

\def\tenpoint{\def\rm{\fam0\tenrm}
\textfont0=\tenrm \scriptfont0=\sevenrm \scriptscriptfont0=\fiverm
\textfont1=\teni \scriptfont1=\seveni \scriptscriptfont1=\fivei
\textfont2=\tensy \scriptfont2=\sevensy \scriptscriptfont2=\fivesy
\textfont3=\tenex \scriptfont3=\tenex \scriptscriptfont3=\tenex
\def\it{\fam\itfam\tenit} \textfont\itfam=\tenit
\def\sl{\fam\slfam\tensl} \textfont\slfam=\tensl
\def\bf{\fam\bffam\tenbf} \textfont\bffam=\tenbf
\scriptfont\bffam=\sevenbf \scriptscriptfont\bffam=\fivebf
\def\tt{\fam\ttfam\tentt} \textfont\ttfam=\tentt
\def\ss{\fam\ssfam\tenss} \textfont\ssfam=\tenss
\tt \ttglue=.5em plus .25em minus .15em
\normalbaselineskip=12pt
\abovedisplayskip=12pt plus 3pt minus 9pt
\belowdisplayskip=12pt plus 3pt minus 9pt
\abovedisplayshortskip=0pt plus 3pt
\belowdisplayshortskip=7pt plus 3pt minus 4pt
\parskip=0.0pt plus 1.0pt
\setbox\strutbox=\hbox{\vrule height8.5pt depth3.5pt width0pt}
\let\sc=\eightrm
\let\big=\tenbig \normalbaselines\rm}
\def\tenbig#1{{\hbox{$\left#1\vbox to8.5pt{}\right.\n@space$}}}
\let\rawfootnote=\footnote \def\footnote#1#2{{\rm\parskip=0pt\rawfootnote{#1}
{#2\hfill\vrule height 0pt depth 6pt width 0pt}}}

\def\tenfoot{\tenpoint\hskip-\parindent\hskip-.1cm}

\overfullrule=0pt
\twelvepoint
\def\sbullet{\raise.2em\hbox{$\scriptscriptstyle\bullet$}}
\nofirstpagenotwelve
\hsize=16.5 truecm
\baselineskip 15pt

\def\ft#1#2{{\textstyle{{#1}\over{#2}}}}

\def\ket#1{\big| #1\big\rangle}
\def\bra#1{\big\langle #1\big|}
\def\braket#1#2{\big\langle #1\big| #2\big\rangle}
\def\dum{{\phantom{X}}}
\def\m{{\rm -}}
\def\p{{\rm +}}

\def\phys{\big|{\rm phys}\big\rangle}

\def\del{\partial}

\def\.{\,\,,\,\,}

\oneandahalfspace
\rightline{CTP TAMU--71/92}
\rightline{hep-th/9211051}
\rightline{November 1992}

\vskip 2truecm
\centerline{\bf Physical States in the $W_3$ String}
\vskip 1.5truecm
\centerline{C.N. Pope\footnote{$^\star$}{\tenfoot Supported in part by the
U.S. Department of Energy, under
grant DE-FG05-91ER40633.}\footnote{}{\tenfoot Contribution to the
proceedings of the 16'th Johns Hopkins Workshop on Current Problems in
Particle \nl \phantom{xxxi} Theory, Gothenburg, Sweden, June 1992.}}

\vskip 1.5truecm

\centerline{\it Center for Theoretical Physics, Texas A\&M University,}
\centerline{\it College Station, TX 77843--4242, USA.}

\vskip 1.5truecm
\AB\singlespace
      We present a review of some of the recent developments in the study of
the $W_3$ string.  One of the interesting features of the theory is that the
physical spectrum includes states with non-standard ghost structure, such as
excitations of the ghost fields, both for discrete-momentum and
continuous-momentum states.
\AE\oneandahalfspace

\np
\noindent
{\bf 1. Introduction}
\bigskip

      The ordinary bosonic string is described by the two-dimensional theory of
matter conformally coupled to gravity.  Quantisation of the two-dimensional
metric tensor requires the fixing of gauge symmetries, and the introduction of
associated Fadeev-Popov ghosts.  The local holomorphic and anti-holomorphic
Virasoro symmetries that remain after fixing the conformal gauge are in
general broken by anomalies at the quantum level.  The anomaly manifests itself
as a central extension of the Virasoro algebra, with the ghosts giving a
contribution of $-26$ to the central charge.  This can be cancelled by choosing
a conformal matter system that gives an equal and opposite central charge.  The
simplest choice is to take 26 free scalars $X^\mu$, with $T=-\ft12\del X_\mu
\del X^\mu$, giving rise to the usual critical 26-dimensional string.  Other
choices are also possible.  In particular, one can realise the Virasoro
algebra with 2 scalars $\varphi$ and $X$, one of which, say $\varphi$, has a
background charge; $T=-\ft12 (\del X)^2-\ft12(\del\varphi)^2 -Q \del^2\varphi$,
with $Q^2=2$. The resulting theory, commonly called the non-critical string, is
still a critical theory in the sense that the matter system has central charge
$+26$, as it must in order to cancel the anomaly.  The $\varphi$ field in
the two-dimensional theory is usually interpreted as a Liouville field.  The
two-dimensional string theory has received much attention recently.  One of
the fascinating features of the theory is that its spectrum of physical
states is much richer and more subtle than one might at first have supposed.

     The physical states in string theory can most
elegantly be described by using the BRST formalism.  The BRST operator $Q_B$ is
given by $$
Q_B=\oint dz c(z)\Big(T(z) + \ft12 T^{\rm gh}(z)\Big),\eqno(1)
$$
where we focus for now just on the holomorphic sector.  Here $T(z)$ is the
energy-momentum tensor for the matter, with central charge 26, and $T^{\rm
gh}$ is the ghost energy-momentum tensor,
$$
T^{\rm gh}= -2b\, \del c- \del b\, c,\eqno(2)
$$
where $c$ and $b$ are the ghost and antighost fields arising from the gauge
fixing for the two-dimensional metric.  The BRST operator is nilpotent,
$Q_B^2=0$, by virtue of the fact that $T$ has central charge 26.

     Physical states are defined to be states in the cohomology of
$Q_B$.  That is to say, a state $\ket{\chi}$ is physical if it satisfies the
physical-state condition
$$
Q_B \ket{\chi}=0,\eqno(3)
$$
and if in addition it is non-trivial, in other words
$$
\ket{\chi}\ne Q_B \ket{\psi}\eqno(4)
$$
for any state $\ket{\psi}$.  In the case of the 26-dimensional string, the
physical states have the ``standard'' form
$$
\ket{\chi}=\ket{\rm phys}\otimes\ket{\m},\eqno(5)
$$
where $\ket{\rm phys}$ involves operators built only from the matter fields,
and $\ket{\m}$ is the ghost vacuum
$$
\ket{\m}\equiv c_1\ket{0},\eqno(6)
$$
obtained from the $SL(2,C)$-invariant vacuum $\ket{0}$ which satisfies
$$
\eqalign{
c_n\ket{0}&=0,\qquad n\ge 2,\cr
b_n\ket{0}&=0,\qquad n\ge -1.\cr}\eqno(7)
$$
For states of the standard form (5), it follows from (1) that the
physical-state condition (3) becomes
$$
\eqalign{
(L_0-1)\ket{\rm phys}&=0,\cr
L_n\ket{\rm phys}&=0,\qquad n\ge1,\cr}\eqno(8)
$$
where $L_n$ are the Laurent modes of $T(z)$.  These are the familiar
physical-state conditions for the 26-dimensional bosonic string.

     To give a full description of the physical states of the string, we
should also consider states of the form
$$
\ket{\chi}=\ket{\rm phys}\otimes\ket{\p},\eqno(9)
$$
where $\ket{\p}\equiv c_0\ket{\m}$.  The reason for this is that a state (5)
actually has zero inner product with itself.  To see this, note that
$c_0\ket{\p}=0$ since the modes $c_n$ anticommute, and similarly
$\ket{\m}=b_0\ket{\p}$ so $b_0\ket{\m}=0$.  Thus we
have $\braket{-}{-}=\bra{+}b_0\ket{\m}$, which therefore vanishes.  In fact the
fundamental non-zero inner product in the ghost sector is
$\braket{+}{-}=\braket{-}{+}=1$.  Thus we must consider states of both forms
(5)
and (9) in order to get non-vanishing inner products.  If we assign ghost
number
$G=0$ to $\ket{\m}$, so $\ket{0}$ has $G=-1$, then $\ket{\p}$ has $G=1$.  In
general, a state with ghost number $G$ will have non-zero inner product with
some state at ghost number $1-G$, which we may think of as its ``conjugate''
state. The inner product on the Hilbert space of states will then be
off-diagonal, leading to an indefinite-signature metric.  To circumvent this we
then truncate out half of the states, by identifying states and their
conjugates.  For the physical states of the 26-dimensional string, this means
identifying states of the forms (5) and (9).  One may think of the standard
physical states of the 26-dimensional string as lying at ghost number $G=0$,
with their conjugates at ghost number $G=1$.

     In the two-dimensional string, things are rather different.  Naively,
one might expect that there should be no physical states ({\it i.e.}\ having
non-zero norm) at all, apart from the tachyons.  This is because all the
excited states of the $d$-dimensional string describe gauge fields, with
gauge invariances that imply that the physical degrees of freedom are
confined to the $(d-2)$-dimensional transverse space.  Thus if $d=2$, one
might think that nothing would survive in the excited spectrum.  As is now
well known, this naive expectation is incorrect.  It turns out that the
analysis in two dimensions has to be handled more carefully, and whilst it is
true for generic values of the on-shell momenta that there are no
positive-norm excited states, there are special cases that arise at specific
values of the on-shell momenta [1].  These are known as discrete states.  The
first examples were of the ``standard'' form (5).  Subsequently, it was
found that there are more general kinds of discrete state in the
two-dimensional string, which do not have the standard ghost structure.  Rather
than having the form (5), they instead involve excitations of the ghost and
antighost operators as well as the matter operators [2].  By considering all
possible states that one could build in this way, it is easy to see that at
a given level number $\ell$, the range of ghost numbers $G$ that could
conceivably occur is $\Big[\ft12 -\ft12 \sqrt{8\ell+1}\Big]\le G\le
\Big[\ft12 +\ft12 \sqrt{8\ell+1}\Big]$, where $\Big[x\Big]$ denotes the
integer part of $x$.  In fact it turns out that the physical states of the
two-dimensional string occupy a band within this allowed range, namely from
$G=-1$ to $G=2$.  (This is wider than the band $G=0$ to $G=1$ for the
continuous states of the 26-dimensional string.)  The states themselves lie in
the range $-1\le G\le 1$, and their conjugates lie in the range $0\le G\le
2$.

     Because of the background charge $Q$ in the $\varphi$ direction, which
can be viewed as an insertion of momentum $-2iQ$ in the $\varphi$ direction at
infinity, momentum conservation in correlation functions is modified.  We
may focus just on the $\varphi$ direction to illustrate the point.  The
momentum operator $\alpha_0$ is not hermitean, but instead satisfies
$\alpha_0^\dagger=\alpha_0-2iQ$.   Thus in the two-point function we have
$\bra{p'}\alpha_0\ket{p}=p\braket{p'}{p}=({p'}^*+2iQ) \braket{p'}{p}$, so
$\braket{p'}{p}$ vanishes unless $p={p'}^* +2iQ$.  This leads to a situation
that is  analogous to that described above for the ghost vacua, in that one now
has  an off-diagonal inner product that leads to an indefinite-signature
metric.   Again the remedy is to halve the dimension of the Hilbert space, by
identifying states and their conjugates.  If we have a
physical state with ghost number $G$ and momentum $\vec p$, then its
conjugate, with which it has a non-zero inner product, has ghost number
$1-G$ and momentum $\vec p^*+(2iQ,0)$.

     The simplest example of a non-standard discrete
physical state is the $SL(2,C)$ vacuum $\ket{0}$.  This may be written as
$b_{-1}\ket{\m}$, since $\ket{\m}=c_1\ket{0}$; it is manifestly invariant
under $Q_B$.  Thus we see that $\ket{0}$ is a level $\ell=1$ physical state,
with ghost number $G=-1$, and momentum $\vec p=0$.  Its conjugate is the
state $\bra{\vec p,0}c_{-1}c_0 c_1$, where $\vec p=(2iQ,0)$.  In fact the
$SL(2,C)$ vacuum $\ket{0}$ is a physical state even in the 26-dimensional
string; it appears to be the only discrete state in that case.

     Discrete states occur at all higher levels in the two-dimensional
string.  Especially interesting are the examples at level $\ell=2$ and
$G=-1$.  They can be described in terms of operators $x$ and $y$ acting on
the $SL(2,C)$ vacuum, with [2]
$$
\eqalign{
x&=\Big( b\, c +\ft12 Q (\del\varphi+i\del X)\Big)
e^{\ft12 Q (\varphi- X)},\cr
y&=\Big( b\, c +\ft12 Q (\del\varphi-i\del X)\Big)
e^{\ft12 Q (\varphi+ X)},\cr}\eqno(10)
$$
with $Q^2=2$.   These operators $x$ and $y$ have conformal spin 0, and ghost
number $G=0$.  By taking products of the form $x^m y^n$, for $m$ and $n$
integer, one can obtain operators that generate higher-level $G=-1$ states
[2].  In fact, it can be shown that {\it all} higher-level $G=-1$ states are
obtained in this way [2,3,4].

     Finally, to fill out the complete spectrum of discrete states in the
two-dimensional string, we note that the operators $a_\varphi\equiv [Q_B,
\varphi]=c\del\varphi -Q\del c$ and $a_X^\dum\equiv [Q_B, X]=c\del X$ , being
BRST commutators, map solutions of the physical-state condition (3) into
new solutions [5].  One might think that these new solutions would be
trivial, in the sense that they could be written as $Q_B$ acting on some
state.  This is, however, not the case, since $\varphi$ and $X$ are not
themselves primary fields.  The situation is closely analogous to writing
the field strength of a Dirac monopole as the exterior derivative of a
potential that has a string singularity.  Since $a_\varphi$ and $a_X^\dum$ each
have ghost charge 1, it follows that by starting with a suitable discrete
state $\ket{G}$ at ghost number $G$, we can create two new discrete states
$a_\varphi\ket{G}$ and $a_X\ket{G}$ at ghost number $G+1$, and one new
discrete state $a_\varphi a_X^\dum\ket{G}$ at ghost number $G+2$.  One can only
apply each of $a_\varphi$ and $a_X$ once, since a repeated application of
the same operator will give a BRST-trivial state.  The situation is
reminiscent of an $N=2$ supermultiplet.

     The upshot of the above considerations is that one obtains all $G=-1$
discrete states of the two-dimensional string by taking arbitrary integer
powers $x^m y^n$ of the ``ground-ring'' generators $x$ and $y$ acting on the
$SL(2,C)$ vacuum.  For each such state, one can then fill out a quartet of
states, with  two at $G=0$ obtained by acting with $a_\varphi$ or
$a_X^\dum$, and one at $G=1$ obtained by acting with both $a_\varphi$ and
$a_X^\dum$.  There will then be a conjugate quartet too, with ghost numbers
$0\le G\le 2$, and momentum conjugate to that of the original quartet.

\bigskip\bigskip
\noindent{\bf 2. The $W_3$ String}
\bigskip

     The essential ingredient for the construction of the bosonic string was
the
existence of a nilpotent BRST operator for the Virasoro algebra.  This has a
Lagrangian interpretation as the existence of an anomaly-free worldsheet
theory of conformal matter coupled to two-dimensional gravity.  We can
now apply similar considerations to the $W_3$ extended conformal algebra.
This is generated by two currents; the spin-2 energy-momentum tensor
$T(z)$, and a spin-3 primary current $W(z)$.   The algebra takes the form
[6]
$$
\eqalign{
T(z) T(w)& \sim {\partial T(w)
\over z-w} +{2 T\over (z-w)^2} +{c/2\over
(z-w)^4},\cr
T(z) W(w)&\sim {\partial W
\over z-w} +{3 W\over (z-w)^2},\cr
W(z)W(w)& \sim {16\over 22+5c}\Big({\partial \Lambda\over
z-w}+{2\Lambda\over (z-w)^2}\Big) +{c/3 \over(z-w)^6}\cr
&\ \ +\ft1{15}\Big({\partial^3 T\over z-w} +\ft92{\partial^2 T\over (z-w)^2}
+15{\partial T\over (z-w)^3} +30{T\over
(z-w)^4}\Big),\cr} \eqno(11)
$$
where $\Lambda=(TT)-\ft3{10}\del^2 T$, with the normal ordering of the product
of operators defined by $(AB)(z)={1\over 2\pi i}\oint dz A(z)B(w)/(z-w)$.

     The BRST operator has the form
$$
Q_B=\oint dz \Big(c(T+\ft12 T_{\rm gh})+\gamma(W+\ft12
W_{\rm gh})\Big), \eqno(12)
$$
and is nilpotent provided that the matter currents $T$ and $W$ satisfy (11)
with central charge $c=100$, and the ghost currents are chosen to be
$$
\eqalignno{
T_{\rm gh}&=-2b\,\partial c-\partial
b\, c-3\beta\, \partial\gamma-2\partial\beta\, \gamma\ , &(13)\cr
W_{\rm gh}&=-\partial\beta\,
c-3\beta\, \partial c-\ft8{261}\big[\partial(b\, \gamma\,  T)+b\,
\partial\gamma \, T\big]\cr
&\ \ +\ft{25}{6\cdot261}\Big(2\gamma\, \partial^3b+9\partial\gamma\,
\partial^2b +15\partial^2\gamma\,\partial b+10\partial^3\gamma\,
b\Big) ,&(14)\cr}
$$
where the ghost-antighost pairs ($c$,$b$) and ($\gamma$,$\beta$) correspond
respectively to the $T$ and $W$ generators [7].  Thus we need a matter
realisation of $W_3$ with central charge 100.  Such realisations can be given
in terms of $n\ge 2$ scalar fields, as follows [8]:
$$
\eqalign{
T&= -\ft12 (\del\varphi_1)^2 - Q \del^2 \varphi_1 +\widetilde T,\cr
W&=-{2i \over \sqrt{261} }\Big( \ft13 (\del\varphi)^3 + Q \del\varphi_1
\del^2\varphi_1 +\ft13 Q^2 \del^3 \varphi_1 + 2\del\varphi_1 \widetilde T
+ Q \del \widetilde T\Big),\cr}\eqno(15)
$$
where $Q^2=\ft{49}{8}$ and $\widetilde T$ is an energy-momentum tensor
with central charge $\ft{51}2$ that commutes with $\varphi_1$.  Since
$\widetilde T$ has a fractional central charge, it cannot be realised simply by
taking free scalar fields.  We can use $d$ scalar fields $X^\mu$, with a
background-charge vector $a_\mu$:
$$
\widetilde T=-\ft12 \del X_\mu \del X^\mu -a_\mu \del^2 X^\mu,\eqno(16)
$$
with $a_\mu$ chosen so that $\ft{51}2=d+12 a_\mu a^\mu$ [9,10].

     As in the case of the ordinary string, we can again look for physical
states that satisfy the condition (3) along  with the condition (4) of
non-triviality. Let us again first consider states of the ``standard'' form
analogous to (5):
$$
\ket{\chi}=\phys\otimes\ket{\m\m},\eqno(17)
$$
where the ghost vacuum is  now given by
$$
\ket{\m\m}=c_1\gamma_1\gamma_2\ket{0}.\eqno(18)
$$
We assign ghost number $G=0$ to $\ket{\m\m}$, so for the $W_3$ string the
$SL(2,C)$ vacuum has $G=-3$.  It satisfies (7), along with
$$
\eqalign{
\gamma_n\ket{0}&=0,\qquad n\ge 3,\cr
\beta_n\ket{0}&=0,\qquad n\ge -2.\cr}\eqno(19)
$$
For states of the form (17), the condition of BRST invariance becomes
$$
\eqalign{
(L_0-4)\phys&=0,\cr
W_0\phys&=0,\cr
L_n\phys=W_n\phys&=0,\qquad n\ge 1.\cr}\eqno(20)
$$
The consequences of these physical-state conditions have been studied in
some detail in various papers [10,11,12,13].  The main features that emerge
are the following. The excited states can be divided into two kinds, namely
those for which there are no excitations in the $\varphi_1$ direction, and
those where $\varphi_1$ is excited too.  For the first kind, we may write
$\phys$ as
$$
\phys=  e^{i\beta\varphi_1(0)}{\phys}_{\rm eff}.\eqno(21)
$$
The physical-state conditions (20) imply that
$$
(\beta-iQ)(\beta-\ft67 iQ)(\beta-\ft87 iQ)=0,\eqno(22)
$$
together with the effective physical-state conditions
$$
\eqalign{
(\widetilde L_0 - \tilde a){\phys}_{\rm eff}&=0,\cr
\widetilde L_n{\phys}_{\rm eff}&=0,\cr}\eqno(23)
$$
where ${\phys}_{\rm eff}$ involves only the $X^\mu$ fields and not
$\varphi_1$.  The value of the effective intercept $\tilde a$ is 1 when
$\beta=\ft67 iQ$ or $\ft87 iQ$, and it equals $\ft{15}{16}$ when $\beta=iQ$.
Thus these states of the $W_3$ string are described by two effective
Virasoro-string spectra, for an effective energy-momentum tensor
$\widetilde T$ with central charge $c=\ft{51}2$ and intercepts $\tilde a=1$
and $\tilde a=\ft{15}{16}$ [10,11,12,13].  The first of these gives a mass
spectrum similar to an ordinary string, with a massless vector at level 1,
whilst the second gives a spectrum of purely massive states.  Explicitly, we
have
$$
\eqalign{
-p^\mu(p_\mu -2i a_\mu)&= 2\ell -2,\qquad \tilde a=1,\cr
-p^\mu(p_\mu -2i a_\mu)&= 2\ell -\ft{15}8,\qquad \tilde a=\ft{15}{16},\cr}
\eqno(24)
$$
where $\ell$ is the level number.

     A crucial property of the states described above is that the values of
allowed $\varphi_1$ momentum, given by (22), occur in conjugate pairs.  As
discussed in the previous section, when a background charge is present the
momentum-conservation law is modified.  Thus for a two-point function to be
non-zero, it  must be that the $\varphi_1$ momenta of the two states must
satisfy $\beta={\beta'}^* +2iQ$.  This is indeed possible for the states
(21), for which $\beta=iQ$ is self-conjugate, whilst $\beta=\ft67 iQ$ and
$\beta=\ft87 iQ$ are conjugates of each other.   Of course we must also
introduce the appropriate notion of conjugation with respect to ghosts as
well, just as we did for the ordinary string.  Here, we do this by defining
$$
\ket{\p\p}=c_0 \gamma_0\ket{\m\m},\eqno(25)
$$
and noting that we therefore have also that $\ket{\m\m}=\beta_0 b_0\ket{\p\p}$,
$c_0\gamma_0\ket{\p\p}=0$ and $b_0\beta_0\ket{\m\m}=0$.  Thus it follows that
$\braket{--}{--}=\braket{++}{++}=0$, whilst we may choose $\braket{++}{--}=
\braket{--}{++}=1$.  The discussion proceeds in parallel to that for the
ordinary string, and we see that here we must introduce the states built
over the $\ket{\p\p}$ ghost vacuum as conjugates of the states (17).  Again
the Hilbert space has off-diagonal inner product, leading to an
indefinite-signature metric, which we remedy by identifying states and
conjugates so as to project onto the positive-definite subspace.

     Now let us consider states that still have the ``standard'' structure
(17), but where now there are excitations in the $\varphi_1$ direction as
well as the $X^\mu$ directions.  It is hard to give a full analysis in this
case, but some partial results have been obtained at low-lying levels.  In
all cases that have been examined, it turns out that the $\varphi_1$
momentum $\beta$ is again frozen to specific values, but these values no
longer occur in conjugate pairs [10,11,12,13]. For example, at level 1 one
finds solutions to the physical-state conditions (20) with
$\beta=\ft{10}7iQ$ and $\beta=\ft{11}7iQ$.  Thus it is not possible to form
any non-zero inner products between such states, and so they are evidently
of zero norm.  One would expect therefore to be able to write them as
BRST-trivial states $Q_B\ket{\psi}$, and indeed this can be done.  This pattern
seems to persist at all higher levels; details are discussed in [13].

     This completes the discussion of ``standard'' states of the form (17) in
the $W_3$ string.  Until recently, it was believed that these would constitute
the full set of physical states in the theory, at least in the case of generic
multi-scalar realisations when the number $d$ of scalars $X^\mu$ exceeds 1.  We
now know, however, that the story is a more complicated one.  For the ordinary
string, it was only in the case of a two-dimensional spacetime that it was
necessary to entertain the possibility of having ``non-standard'' physical
states with ghost as well as matter excitations.  For the $W_3$ string,
however,
it turns out that for any dimension for the spacetime, {\it i.e.}\  for any
number $d\ge 1$ of $X^\mu$ fields, states with non-standard ghost structure
will
play a r\^ole.\footnote{$^*$}{\tenfoot  An example of a discrete state with
non-standard ghost structure in the special case of the two-scalar $W_3$ string
was first exhibited in [14].}   Thus we consider general states built from
creation operators for the $b$, $c$, $\beta$ and $\gamma$ ghosts as well as
the matter fields, acting on tachyonic states of the form $\ket{\vec p,--}$.
Since we have assigned ghost number $G=0$ to the tachyonic state $\ket{\vec
p,--}\equiv e^{i\beta\varphi_1 +i p_\mu X^\mu}\ket{\m\m}$, then it is easy
to see by considering all possible excited states that at a given level
number $\ell$ their ghost numbers can lie in the range
$\Big[1-\sqrt{4\ell+1}\Big] \le G \le \Big[1+\sqrt{4\ell+1}\Big]$ [15].  As
for the case of the ordinary string, it looks as if the physical-state
condition (3), together with the condition of non-triviality (4), singles
out states in a restricted band of ghost numbers within the limits given
above.  The evidence accumulated so far suggests that this band for
non-trivial physical states may lie in the range $-3\le G\le 5$ ({\it i.e.}\
$-3\le G \le 1$ for the states themselves, and $1\le G \le 5$ for their
conjugates).

     Preliminary investigations of the new kinds of physical states for the
$W_3$ string, with ``non-standard'' ghost structure, were carried out in
[15].  The main thrust of this work was a study of physical discrete states
in the two-scalar $W_3$ string ({\it i.e.}\ the case when there is just one
$X^\mu$ coordinate in addition to the field $\varphi_1$).  These could be
viewed as  the direct analogues of the discrete states of the two-scalar
ordinary  string.  We shall describe them in more detail later. However, it
also emerged from this work that there are new physical states with
non-standard ghost structure in the multi-scalar $W_3$ string  too; {\it
i.e.}\ when there is an arbitrary number $d\ge 1$ of additional  coordinates
$X^\mu$.  This is a rather surprising result, which indicates  that the
$W_3$ string theory is very different, and in some sense richer,  than the
ordinary string.  It is not only for discrete states, but also for  true
physical states with continuous on-shell momenta in the multi-scalar  $W_3$
string, that states with non-standard ghost structure appear.  We  shall
illustrate this with an example that was found in [15], namely states  at
level $\ell=1$ and $G=-1$.

     From the general formula given above, one sees that the allowed range
of ghost numbers at level $\ell=1$ is $-1\le G\le 2$.  Let us consider the
most negative ghost number, namely $G=-1$.  Since $Q_B$ has ghost number
$G=1$, it follows that any $G=-1$ state at $\ell=1$ that is annihilated by
$Q_B$ must be BRST non-trivial, since no state $\ket{\psi}$ at $G=-2$ can
exist to allow the existence of a $G=-1$ trivial state $Q_B\ket{\psi}$.  Just
two structures are possible for states at $G=-1$, corresponding to the operator
$$
\Big( c\, \gamma + \alpha\  \gamma\del\gamma\Big) e^{i\beta \varphi_1 + i
p_\mu X^\mu},\eqno(26)
$$
where $\alpha$ is a constant. Requiring that this be annihilated
by the BRST operator leads to the following results.  There are two discrete
states, with momenta
$$
(i\beta,p_\mu)=(-\ft67 Q, 0),\quad (-\ft87 Q,0),\eqno(27)
$$
and two continuous-momentum states, with momenta
$$
(i\beta,p_\mu)=(-\ft37 Q,p_\mu),\quad (-\ft47 Q,p_\mu),\eqno(28)
$$
where for each state the ``spacetime'' momentum $p_\mu$ satisfies
$$
-p^\mu(p_\mu-2i a_\mu)=-1. \eqno(29)
$$
The constant $\alpha$ in (26) is equal to $i/\sqrt{522}$ for the
continuous state with $\beta =\ft47 iQ$ and the discrete state with $\beta
=\ft67 iQ$.  In the other two cases, $\alpha=-i/\sqrt{522}$.

     These level $\ell=1$ states at $G=-1$ provide the first examples of
physical states with non-standard ghost structures in the multi-scalar $W_3$
string.  Just as in the case of the two-dimensional ordinary string, we may
now build ``multiplets'' of further states, at higher ghost numbers, by
acting with the operators $a_\varphi\equiv [Q_B,\varphi_1]$ and $a_{X^\mu}^\dum
\equiv [Q_B,X^\mu]$.  Their form is more complicated than for the ordinary
string; details may be found in [15].  By this means we may obtain from the
$G=-1$ level-1 states discussed above further $\ell=1$ states at higher ghost
numbers including $G=0$.  The $G=-1$ states (26) have excitations
exclusively in the ghost operators.  The ``multiplet'' states that we build
by acting with $a_\varphi$ and $a_{X^\mu}^\dum$ will have matter excitations as
well as ghost excitations.  These will include excitations in the
$\varphi_1$ direction as well as $X^\mu$.  Thus the previous conclusion that
all states with excitations in the $\varphi_1$ direction have zero norm is
seen to be true only for states with the ``standard'' ghost structure (17).
The states that we have just been describing, which were constructed in [15],
all of course have non-zero norm, since they are BRST non-trivial.

     The BRST-non-trivial physical states that we have been describing are
just the first examples of a complex pattern of states.  Work on analysing this
in more detail is in progress [16].

\bigskip\bigskip
\noindent{\bf 3. Discrete states in the two-scalar $W_3$ string}
\bigskip

     The physical states that we have been considering so far have been for
multi-scalar $W_3$ strings.  In view of the experience with ordinary string
theory, where the two-scalar string has its own interest and surprises, it
is natural to wonder what might happen in the analogous situation for the
$W_3$ string.  As discussed in [15], it is not totally clear what the analogue
of the two-scalar string should be in the $W_3$ case.  Since the
physical-state conditions always lead to the ``freezing'' of the momentum in
the $\varphi_1$ direction, one might think that a total of three scalars,
namely $\varphi_1$ plus two more $X^\mu$ coordinates, would be the natural
analogue.  However, the physical states of the $W_3$ string have fewer gauge
symmetries than those of the ordinary string.  For example, we saw in the
last section that there are excited states described by effective Virasoro
strings as in (23), with effective intercept $\tilde a=\ft{15}{16}$.  One
can see from the fact that the level-1 vector is massive that it will have
non-zero norm for arbitrary on-shell momenta in the three-scalar $W_3$
string.  Thus there will be continuous-momentum excited states as well as
discrete states in a three-scalar $W_3$ string.  In order to have a theory
with only discrete-momentum states, it is necessary to consider the
two-scalar $W_3$ string.  Thus we have coordinates $\varphi_1$ and
$X=\varphi_2$, with the background-charge vector $a^\mu$ in (16) becoming
just the one-component quantity $a$, such that $a^2=\ft{49}{24}$.

     In section 1, we saw that the $SL(2,C)$ vacuum in the two-scalar
ordinary string is a discrete physical state at level $\ell=1$ and ghost
number $G=-1$.  In the two-scalar $W_3$ string it is also clear that
$\ket{0}$ is a physical state, in that it satisfies $Q_B\ket{0}=0$ and it
cannot be written as $Q_B\ket{\psi}$ for any $\ket{\psi}$.  In this case, it
is a level $\ell=4$ state with ghost number $G=-3$, since it can be written as
$$
\ket{0}=\beta_{-2}\beta_{-1}b_{-1}\ket{\m\m}.\eqno(30)
$$
In the two-scalar string, the $x$ and $y$ operators (10) that give rise to
$G=-1$ states at level $\ell=2$ played a crucial r\^ole, since they could be
used to construct all higher-level $G=-1$ states by acting with $x^m y^n$ on
the $SL(2,C)$-invariant vacuum.  In [15], a search was undertaken for
analogous operators in the two-scalar $W_3$ string.  It was found that two
such operators exist that correspond to $G=-3$ states at level $\ell=6$.
They have the form
$$
\eqalign{
x&=R_x e^{\ft27 Q \varphi_1},\cr
y&=R_y e^{(\ft17 Q \varphi_1 +\ft37 a \varphi_2)},\cr}\eqno(31)
$$
where the prefactors $R_x$ and $R_y$ involve the 30 different possible
excitation structures for this level and ghost number; $R\sim g_1 b\, \del b
\,\gamma \,\del \gamma+\cdots + g_{30} b\,\gamma (\del \varphi_2)^2$.
Demanding BRST invariance fixes the momenta to be those given in (31), and
determines all but two of the constant coefficients $g_i$ in each case.
Only one of these two parameters represents a BRST-non-trivial state; the
other is a reflection of the fact that there exists a unique state at the
minimum allowed ghost number $G=-4$ at level 6, and so $Q_B$ of this state
gives a trivial BRST-invariant state at $G=-3$.

     Thus we find the two non-trivial operators $x$ and $y$ at level 6 [15].
Like the analogous operators in the two-scalar string, they have spin 0 and
ghost number $G=0$.  One can thus build new such operators by taking
products of powers of $x$ and $y$.  In the string case, one can take
arbitrary integer powers; this is because the $(z-w)$ pole that one gets
from normal ordering $x$ with $x$, $y$ with $y$, or $x$ with $y$ is always
of integer degree.  In the $W_3$ case, however, this is no longer true.  One
can see from (31), and the definitions of the background charges
($Q^2=\ft{49}8$, $a^2=\ft{49}{24}$), that the normal ordering of the
exponentials will give factors $(z-w)^{-1/2}$ for $x x$ and $y y$, and a
factor of $(z-w)^{-1/4}$ for $x y$.  Thus only certain powers of these
operators will be well defined. It turns out that the full set of allowed
powers is given by [15]
$$
x^{4p} y^{4q} \{1,\ x,\ y,\ xy^2,\ x^2 y,\ x^2 y^2\},\eqno(32)
$$
where $p$ and $q$ are arbitrary non-negative integers.  The situation here
is similar to that for the ordinary string in one dimension rather than two,
where again only certain powers of the single $x$-type operator that exists
in that case can arise [15].

     As discussed in [15], there are indications that further spin 0, $G=0$
operators should exist at higher levels, which cannot be built from powers
of $x$ and $y$.  It seems likely that four such operators at level $
\ell=8$, with momenta $(i\beta,ip)=(\ft47 Q, -\ft27 a),\quad (\ft47 Q,
-\ft{12}7 a),\quad (\ft17 Q, a)$ and $(-\ft47 Q, \ft{12}7 a)$, should be
sufficient, together with $x$ and $y$ above, to generate all spin-0, $G=0$
physical operators in the two-scalar $W_3$ string.  Full details of this
will appear elsewhere [17].

\bigskip\bigskip
\noindent{\bf 4. Conclusions}
\bigskip

     We have seen that in the $W_3$ string not only are there
physical states with the ``standard'' ghost structure (17), but also there
are further physical states involving ghost excitations as well as matter
excitations.  The simplest examples, at level $\ell=1$, were found in [15],
and are given in (26)-(29).  It is indeed fortunate that such physical
states occur, since otherwise, in the standard sector (17), only states of
the form (21) with no excitations in the $\varphi_1$ direction arise.  Since
it is the $\varphi_1$ coordinate that is the truly special and distinctive
field of the $W_3$ realisations (it is the only one that does not enter
exclusively {\it via} its energy-momentum tensor), it would otherwise have
been the case that the $W_3$ string spectrum amounted to little more than a
set of two slightly unusual Virasoro string spectra.  Instead, it turns out
that the truly new aspects of the $W_3$ string spectrum occur in the
non-standard ghost sectors.

     Perhaps the most intriguing aspect of the $W_3$ spectrum is that the
physical states with non-standard ghost structure are important in the
multi-scalar case as well as the two-scalar case.  Thus by contrast with the
ordinary string, the non-standard states play a r\^ole for the
continuous-momentum physical states of a multi-scalar $W_3$ string as well
as for the discrete states of a two-scalar $W_3$ string.  The reason for
this may be related to the smaller number of gauge symmetries of the states of
the  $W_3$ string.  The ordinary string has gauge symmetries that restrict the
degrees of freedom of the continuous states in the $d$-dimensional theory to
the $(d-2)$-dimensional transverse space.  Thus, aside from the discrete
states, there exists a physical light-cone gauge.  (In fact the breakdown of
the light-cone gauge occurs precisely at the discrete momenta for which the
denominator $P^+$ becomes zero.)  The gauge symmetries of
the $W_3$ string are fewer, as indicated for example by the presence of a
massive vector in its spectrum, and so there is no notion of a physical
light-cone gauge.

     An interesting way of understanding the necessity of the new physical
states with non-standard ghost structure in the $W_3$ string has been discussed
recently in [18].  In this paper, the authors calculate the scattering
amplitude
for four tachyons in the $\tilde a=\ft{15}{16}$ branch of the ``standard''
states given by (17) and (21).  This involves the introduction of the
conformal-spin $\ft1{16}$ primary field $\sigma$ of the Ising model, in order
to make up the deficit of the spin $\ft{15}{16}$ of the tachyon operators
$e^{ip\cdot X}$ to give spin-1 vertex operators.  In [18], the resulting
scattering amplitude is then expanded in $s$-channel poles, revealing the
existence of resonances with masses that are consistent with neither of the
mass formulae in (24), which are associated with states of ``standard'' ghost
structure.  In fact, the calculation shows that there must be extra
states in the $W_3$ string with masses of the form (28), together with
higher-level excitations.  Thus one sees by looking at the scattering of
``standard'' states that there must also be states of non-standard ghost
structure in the physical spectrum.

     Finally, we remark that recently a new kind of BRST operator for $W_3$ has
been constructed, starting from two independent and mutually-commuting copies
of the $W_3$ algebra [19].  This may be relevant to a description of
non-critical $W_3$ strings.  A field theory giving rise to this BRST
operator has been constructed [20].  Many issues analogous to those discussed
in this paper could be investigated for this case.

\bigskip\bigskip
\centerline{\bf Acknowledgments}
\bigskip

     I am very grateful to my collaborators in the work described in this
review, namely Hong Lu, Bengt Nilsson, Larry Romans, Stany Schrans, Ergin
Sezgin, Kelly Stelle, Xujing Wang, Peter West and Kaiwen Xu.

\bigskip\bigskip

\noindent{\bf Note Added}
\bigskip

     After this paper was completed a paper considering related issues for the
new BRST operator of [19] appeared [21].

\np

\centerline{\bf REFERENCES}
\frenchspacing
\bigskip

\item{[1]}A.M. Polyakov, {\sl Mod. Phys. Lett.} {\bf A6} (1991) 635.

\item{[2]}E. Witten, {\sl Nucl. Phys.} {\bf B373} (1992) 187.

\item{[3]}B.H.  Lian and G.J. Zuckermann, {\sl Phys. Lett.} {\bf B254}
(1991) 417.

\item{[4]}P. Bouwknegt, J. McCarthy and K. Pilch, {\sl Comm. Math. Phys.}
{\bf 145} (1992) 541.

\item{[5]}E. Witten and B. Zwiebach, {\sl Nucl. Phys.} {\bf B377} (1992) 55.

\item{[6]}A.B. Zamolodchikov, {\sl Teo. Mat. Phys.} {\bf 65} (1985) 347.

\item{[7]}J. Thierry-Mieg, {\sl Phys. Lett.} {\bf B197} (1987) 368.

\item{[8]}L.J. Romans, {\sl Nucl. Phys.} {\bf B352} (1991) 829.

\item{[9]}A. Bilal and J.-L. Gervais, {\sl Nucl. Phys.} {\bf B314}
(1989) 646; {\bf B318} (1989) 579;\nl
S.R. Das, A. Dhar and S.K. Rama, {\sl Mod. Phys. Lett.}
{\bf A6} (1991) 3055;\nl
{\it Physical states and scaling properties of $W$  gravities and $W$
strings}, TIFR/TH/91-20.

\item{[10]}C.N. Pope, L.J. Romans and K.S. Stelle, {\sl Phys. Lett.}
{\bf B268} (1991) 167; {\sl Phys. Lett.} {\bf B269} (1991) 287.

\item{[11]}C.N. Pope, L.J. Romans, E. Sezgin and K.S. Stelle, {\sl Phys.
Lett.}  {\bf B274} (1992) 298.

\item{[12]}H. Lu, C.N. Pope, S. Schrans and K.W. Xu, {\it The Complete Spectrum
of the $W_N$ String},  preprint CTP TAMU-5/92, KUL-TF-92/1, to appear in {\sl
Nucl. Phys. B}.

\item{[13]}H. Lu, B.E.W. Nilsson, C.N. Pope, K.S. Stelle and P.C. West,
in preparation.

\item{[14]}S.K. Rama, {\sl Mod. Phys. Lett.} {\bf A6} (1991) 3531.

\item{[15]}C.N. Pope, E. Sezgin, K.S. Stelle and X.J. Wang, {\it Discrete
States in the $W_3$ String}, preprint, CTP TAMU-64/92.

\item{[16]}H. Lu, C.N. Pope and P.C. West, in progress.

\item{[17]}C.N. Pope and X.J. Wang, in progress.

\item{[18]}M.D. Freeman and P.C. West, {\it $W_3$ String Scattering},
preprint, KCL-TH-92-4.

\item{[19]}M. Bershadsky, W. Lerche, D. Nemeschansky and N.P. Warner, {\it A
BRST Operator for Non-critical $W$ Strings}, preprint, CERN-TH.6582/92.

\item{[20]}E. Bergshoeff, A. Sevrin and X. Shen, {\it A Derivation of the BRST
Operator for Non-critical Strings}, preprint, CERN-TH.6647/92.

\item{[21]}M. Bershadsky, W. Lerche, D. Nemeschansky and N.P. Warner, {\it
Extended $N=2$ Superconformal Structure of Gravity and $W$-Gravity Coupled to
Matter}, preprint, CERN-TH.6694/92.

\end